# Classical Möbius-Ring Resonators Exhibit Fermion-Boson Rotational Symmetry

**Douglas J. Ballon**

Department of Physiology, Biophysics, and Systems Biology

Weill Cornell Graduate School of Medical Sciences

1300 York Avenue, Box 234

New York, NY 10021

Corresponding author.

Email: dballon@med.cornell.edu

Phone: 212 746 5679

Fax: 212 746 6681

**Henning U. Voss**

Department of Radiology

Weill Cornell Medical College

Citigroup Biomedical Imaging Center

516 E 72$^{nd}$ Street

New York, NY 10021

Email: hev2006@med.cornell.edu

**Abstract**. The behavior of coupled harmonic oscillators in systems with specified boundary conditions is typically characterized by resonances whose frequency spectra represent harmonics according to properties of the individual oscillators, the interactions between them, and the overall symmetry of the system. Here it is demonstrated that classical one- and two-dimensional radiofrequency resonators constrained to a Möbius topology are the formal partners of cylindrical ring resonators, and specifically give rise to half-integral harmonic excitations that are orthogonal to the integral excitations of a ring. In particular, the half-integral harmonics are formally invariant under rotations at a minimum of $4\pi$ rather than $2\pi$ radians, in analogy to the rotational symmetry of fermions in quantum mechanics. The results offer a pathway for discovery in other physical systems as well as the design of novel materials and electronic instrumentation.

Since its discovery in 1858, the Möbius strip has evolved from a mathematical and artistic curiosity to a recent topological target in the design of materials [1], molecules [2-6], nanostructures [7,8] and electronic and microwave devices [9-13]. The purpose of this letter is to point out that certain classical physical systems embedded in the Möbius band topology are formal complements to the identical systems defined on a cylindrical ring. The two systems are related by a single topological transformation that can be described by a simple boundary condition, yet exhibit markedly different behavior. As an example, we discuss the rotational properties of eigenfunctions describing classical one- and two-dimensional radiofrequency resonator pairs. The results suggest that classical analogs of phenomena previously associated only with half-integral spin in the quantum mechanical regime may be more common than previously believed.

Radiofrequency resonators are a special class of transmission lines of finite extent with well-defined boundary conditions. They have been used as electrical filters since the earliest days of signal transmission [14], and more recently have found wide application in areas such as magnetic resonance imaging [15]. Of particular interest here is a one-dimensional radiofrequency resonator ring that can be represented schematically by the simple low-pass filter circuit of Fig. 1a. In solving the problem for the electric currents on the resonator, we require a periodic boundary condition of the form

$$I_{n+N} = I_n , \qquad (1)$$

where $I_n$ represents the electric current amplitude around the *n*th closed loop on the periodic ladder structure of *N* elements. Eq. (1) is merely a statement of invariance of the solutions under a $2\pi$ rotation. A resonator whose eigenfunctions satisfy Eq. (1) is shown in Fig. 1b. Its behavior under an applied broadband alternating voltage is well known, and consists of discrete normal modes of oscillation whose frequencies depend upon the inductance and capacitance of each element as well as the mutual inductance between them [16].

Boundary conditions of the general form shown in Eq. (1) apply to virtually any discrete classical periodic system. More specific to the problem of a radiofrequency resonator ring is the fact that, as defined, the *I$_n$* are conserved quantities with a definite handedness. This formally defines a distinct inner and outer surface for the ring of Fig. 1b.

Next, consider a topological transformation resulting in a Möbius resonator of the form shown in Fig. 1c [17]. To describe its behavior, a twisted boundary condition is necessary:

$$I_{n+N} = -I_n \,. \tag{2}$$

That is, a simple topological transformation on the resonator ring results in a sign reversal of current amplitude upon a $2\pi$ rotation of the solutions, and a $4\pi$ rotation is now required for invariance of the eigenfunctions.

Before explicitly deriving the consequences of Eqs. (1) and (2), we note that the theory is in principle not restricted to radiofrequency resonators, but the following discussion will remain so confined since it will lead directly to a comparison with experiments carried out on a simple prototype. In Fig. 1a, a mutual inductance $M$ is defined between nearest neighbor elements only, represented by the coefficient $\kappa$, where $M = 2\kappa L$. Resistance is not essential for this discussion and therefore neglected, and only the non-radiative, or near-field regime will be treated. That is, we will assume that the electromagnetic wavelength is much larger than the linear dimensions of the network. Finally, it is assumed that the circuit is driven by a voltage that varies sinusoidally with time across any one of the capacitors or inductors. With these constraints, the application of Maxwell's Equations leads to Kirchhoff's voltage relation for the $n$th element,

$$\left(\omega^2 - \frac{1}{LC}\right) I_n + \left(\frac{1}{2LC} - \kappa\omega^2\right)(I_{n+1} + I_{n-1}) = 0 \,. \tag{3}$$

Eq. (3) is simply the one-dimensional non-dissipative wave equation on the circuit, with a discrete rather than continuous spatial variable.

For the ring of Fig. 1b with the periodic boundary condition given in Eq. (1), solutions are of the usual form

$$I_n = A \exp\left(\frac{i2\pi n\Omega}{N}\right) + B \exp\left(-\frac{i2\pi n\Omega}{N}\right). \tag{4}$$

Here $\Omega$ is an integer specifying the normal mode. Substitution of Eq. (4) into Eq. (3) yields the dispersion relation for the allowed frequency spectrum:

$$\omega^2 = \frac{2\sin^2\left(\frac{\pi\Omega}{N}\right)}{LC\left(1-2\kappa\cos\left(\frac{2\pi\Omega}{N}\right)\right)}. \tag{5}$$

For an *N* element structure where *N* is even, there are *N*-1 eigenvalues, including (*N*-2)/2 degenerate doublets and one singlet.

Next, consider the twisted boundary condition of Eq. (2), and note that the eigenfunctions satisfying the condition are of the same form as Eq. (4) provided that the mode indices are given half-integral values: $\Omega$ = 1/2, 3/2, 5/2,…(*N*-1)/2 relative to a ring consisting of identical components. The dispersion relation is therefore *identical* to Eq. (5); however, the wavevectors are shifted by

$$\Delta k = -\pi/N . \tag{6}$$

The two distinct topological entities of Fig. 1b,c can thus be viewed formally as a complementary pair related by a single transformation. Their description naturally divides into half-integral and integral normal mode indices, and the rotational properties of the eigenfunctions describing the current amplitudes have a simple analogy to the transformation of half-integral (fermion) and integral (boson) spin wavefunctions under rotation. More formally, the half-integral mode indices guarantee that the current amplitude eigenfunctions $I_n$ transform as spinors on the twisted structure.

Formally orthogonal degenerate doublets are well known in the ring resonator, and are realized in practice by driving the ring at circuit elements spaced $\pi/2$ radians apart. They are also supported in the Möbius resonator according to Eq. (4). However, in this case they are excited in a non-intuitive way by driving the circuit at elements spaced $\pi$ rather than the conventional $\pi/2$ radians apart.

Eigenvalues and eigenfunctions for a Möbius-ring pair constructed in the laboratory are shown in Fig. 2. Normal mode spectra for the eight element structures are also presented. The eight element ring resonator of Fig. 1b was constructed using 16.0 ± 0.8 pF porcelain chip capacitors soldered to adhesive backed copper tape of width 6 mm and thickness equal to 40 μm, which was in turn was fixed to a strip of Teflon with a width of 4.5 cm and thickness equal to 1 mm. The finished resonator had a diameter of 22 cm and a width equal to 4 cm. The distance between capacitive elements was approximately 9 cm. The Möbius resonator of Fig. 1c was formed directly from the ring topology by cutting the two copper tape end rings and effecting a topological transformation on the Teflon substrate via a half twist, then re-soldering the copper leads. Normal modes were measured via voltage reflection from a return loss bridge circuit and a network analyzer. Excitation frequencies were swept from 25-275 MHz in order to span the entire band of resonances (eigenvalues). Nominally degenerate pairs were observed due either to the slight variation in lumped elements in the circuit, or by driving the structures successively at *N/2* or *N/4* elements apart for the Möbius and ring resonators respectively. An alternative method for distinguishing degenerate pairs was to douse the resonator with liquid nitrogen. The resulting reduction in linewidths provided the necessary spectral resolution for visualization of the doublets. Eigenfunctions were mapped using a continuous wave excitation frequency positioned at each eigenvalue. A total of seven doublets and one singlet were observed.

The data for both structures are plotted together along with predicted dispersion relations for several values of the mutual inductance coupling constant $\kappa$. The best fit to the data yielded $\kappa = 0.07$. As predicted, the eigenvalues of the Möbius resonator fall precisely at half-integral values of the wavevectors when compared to its cylindrical partner.

By inspection of Eqs. (1) and (2), it is evident that there is no additional structure associated with the Möbius-ring pair, since a second topological half-twist transformation on the Möbius resonator leads back to the boundary condition of Eq. (1). That is, for one-dimensional resonators, it is evident that only integral and half-integral excitations are allowed, and other sub-integral excitations are excluded. However, since the Möbius resonator has a definite handedness, there are two topologically distinct Möbius resonators of opposite chirality with identical eigenfunctions and eigenvalues.

Note that the one-dimensional Möbius-ring resonator pair may be one of the simplest possible classical manifestations of sub-integral harmonic behavior, as defined by the symmetry of the eigenfunctions under rotation. For example, although there are numerous examples of isomorphism between electrical and mechanical systems in nature, and in particular Eq. (3) with $\kappa = 0$ is isomorphic to a one-dimensional system of coupled mechanical oscillators, the boundary condition of Eq. (2) is the point of departure for similarities between the two domains as there is no simple analog of spinor behavior in the mechanical case.

From the dispersion relation of Eq. (5), it is clear that the eigenvalues are a relatively insensitive function of the mutual inductance, and thus the deformation of classical Möbius strip substrates that was recently demonstrated [8] does not appreciably affect the behavior of the Möbius resonator. In fact, for our prototype, if $\Omega = N/4$ the resonance frequency is independent of $\kappa$.

The eigenfunctions of the Möbius resonator form an orthogonal basis set, which is also the case for the ring. It is also true that all eigenfunctions of the pair are mutually orthogonal. This presents an interesting possibility for the design of metamaterials [18]. For example, if the material is composed of an array of either Möbius or ring resonators as its basis, complement resonators could in principle be introduced with minimal electromagnetic interactions.

Observation of $4\pi$ rotational symmetry in systems described by classical eigenvalue problems has thus far been rare in nature. However, twisted boundary conditions with a generalized form of Eq. (2) have recently been discussed for systems of Heisenberg spin rings [19], leading to shifts in wavevectors of magnitude $\Delta k = \Lambda/N$ (compare to Eq. (6)), where $\Lambda$ is the solid angle of the external field subtended across the ring and $N$ is the number of spin sites.

The investigation of analogous rotational symmetry in two-dimensional systems is in principle straightforward using radiofrequency resonators as models [20,21]. There are several interesting possible topologies, including the Klein bottle and twisted toroidal resonators. As a simple extension of the theory which illustrates some characteristics of classical two-dimensional twisted symmetry, we describe a two-dimensional high-pass Möbius resonator, or twisted periodic drum, and compare the results to experimental findings. The schematic resonator shown in the inset to Figure 3a is a 2 × 8 element twisted ladder network. If we retain the assumption of nearest neighbor coupling but allow rectangular rather than square meshes, the eigenvalue problem can be defined as a straightforward extension of Eq. (3). However, the boundary conditions on the eigenfunctions are now

$$I_{0,n} = I_{M+1,n} = 0; \quad I_{M+1-m,n+N} = -I_{m,n}. \tag{7}$$

With eigenfunctions of the form

$$I_{m,n} = \sin\left(\frac{\pi m \Omega}{M+1}\right)\left[A \exp\left(\frac{i2\pi n\Gamma}{N}\right) + B \exp\left(-\frac{i2\pi n\Gamma}{N}\right)\right], \tag{8}$$

it follows that two excitation bands are allowed on the structure, as shown in Figure 3b. The higher frequency band is given by $\Omega = 1, \Gamma = (1/2, 3/2, 5/2, 7/2)$, while the lower frequency band has $\Omega = 2, \Gamma = (0, 1, 2, 3, 4)$. Thus, in two dimensions, eigenfunctions are allowed with both $2\pi$ and $4\pi$ rotational symmetry on the same structure. Note also that in a high-pass configuration, the most homogeneous eigenfunctions appear at the highest frequency. If the results are compared to those of a 2 × 8 ring resonator, it can be shown that the high frequency band of the ring has $\Omega = 1, \Gamma = (0, 1, 2, 3, 4)$ in analogy to the Möbius-ring correspondence for the one-dimensional case, while the low frequency band has identical eigenfunctions and eigenvalues for the two structures. This behavior was verified on a 2 × 8 element resonator constructed on a Teflon band 80 cm in length and 15 cm wide, and the results plotted in Figure 3b. In this case no free parameters were used to fit the data. Rather, the single element resonant frequency and the long- and short-sided inductive coupling constants were measured on single and two-loop resonators respectively, and these values used in the theoretical calculations.

One potential application of the above results is in the design of electronic musical instruments [22]. If we use the one-dimensional prototype as an example, note that the dispersion relation is nearly linear for $\Omega < N/4$. A transformation from the ring to the Möbius structure lowers the fundamental frequency by one octave and higher frequencies in proportion to the mode numbers. In two dimensions, the normal modes are similar to those of a snare drum shell [23]. In either case there is a choice of a conventional or inverted frequency scale depending upon whether a low-pass or high-pass configuration is used. Using standard methods to shift the radiofrequencies into the audio range, and the fact that an inductively coupled electronic hammer device can be designed to "strike" the resonator at different locations and thereby excite harmonics at variable amplitudes, may offer interesting opportunities for large $N$. An electronic or rapid mechanical switch from the ring to the Möbius topology would offer the possibility for two similar but distinct musical scales on the same device.

In closing we point out that since classical analogs of fermion-boson rotational symmetry are evidently possible at length scales spanning several orders of magnitude, observation of the physics in other systems or applications in device design appear to possible for a range of engineering and physical problems. If only spectroscopy is available to infer the topology, the resolution of the wavevectors is critical, and therefore systems with lower $N$ may be the preferred candidates.

**Acknowledgment**

We thank Eric Aronowitz for assistance in constructing the resonators.

**Figures**

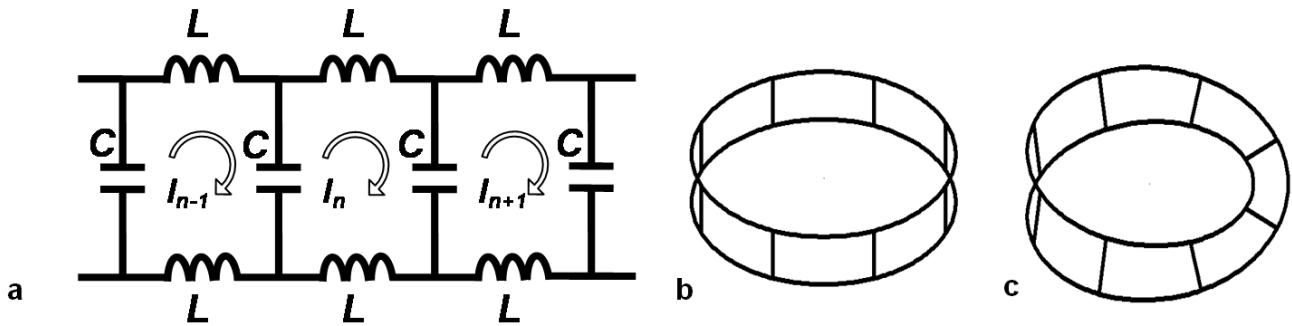

**Figure 1:** The Möbius-ring resonator pair. **(a)** An idealized one-dimensional low-pass ladder network consisting of a series of inductances $L$ and capacitances $C$ along the line. Mutual inductances are considered between the loop elements defined by $I_n$ (see text). There are three options for making a periodic structure from the segment shown. They include the ring resonator in **(b)**, the Möbius resonator in **(c)**, and a Möbius resonator of opposite chirality (not shown). The discrete inductive and capacitive elements in **(b)** and **(c)** are omitted for clarity.

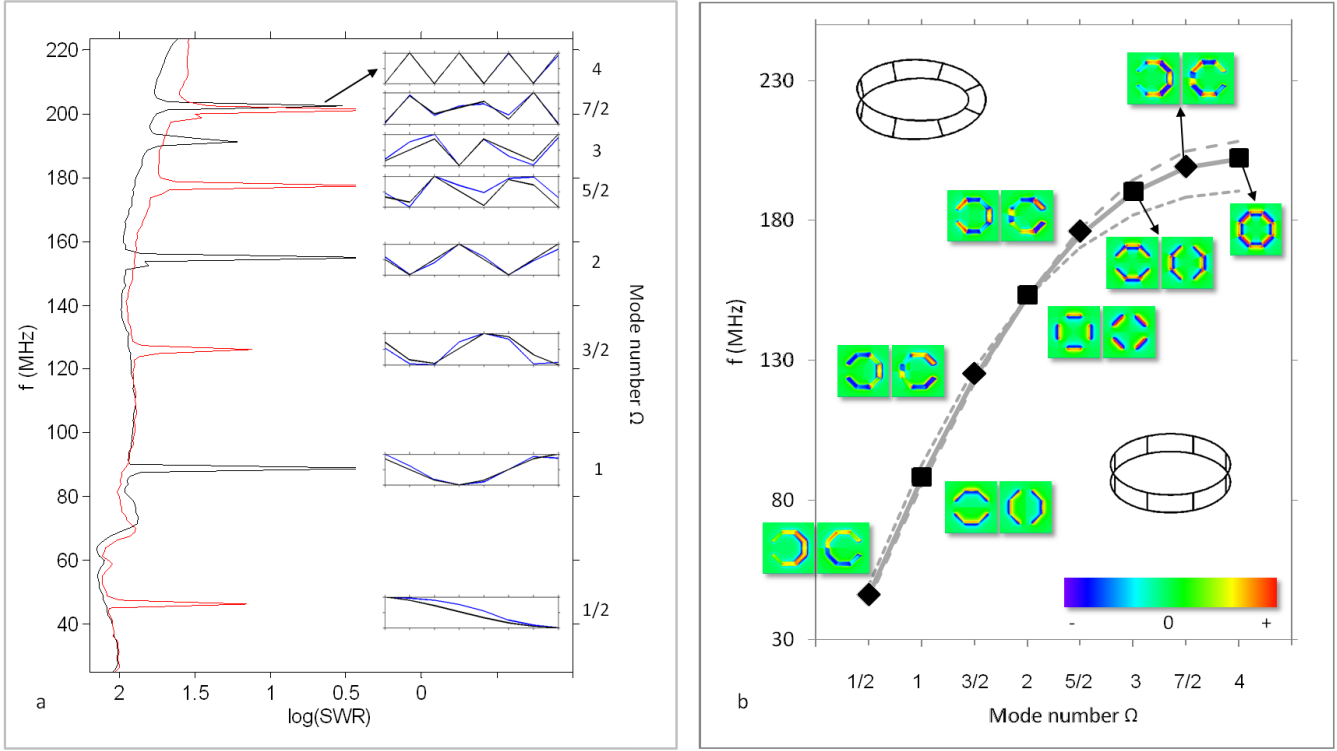

**Figure 2:** Eigenvalues and eigenfunctions for one-dimensional low-pass Möbius-ring resonators. **(a)** Measured spectra (eigenvalues) are shown for the Möbius (red) and ring (black) resonators. The spectral amplitudes are the voltage standing wave ratios observed while driving the resonators inductively at a single location. Using this method all resonant frequencies are visible, but the degenerate doublets are unresolved in most cases (see text for details). The insets show theoretical (black, following Eq. (4)) and measured current amplitudes (blue), respectively. **(b)** If the measured eigenvalues of the Möbius resonator are specifically assigned to half-integral mode indices according to Eq. (5), they lie along the identical theoretical dispersion curve (in bold gray, $\kappa = 0.07$) describing the conventional ring resonator. The dotted curves represent one half (- - -) and twice (---) the observed coupling constant, and reflect a minimal change in the dispersion relation. Inset (color plots): The fifteen predicted eigenfunctions for the eight element Möbius-ring pair represented by Biot-Savart plots of the magnetic field amplitude at each eigenvalue.

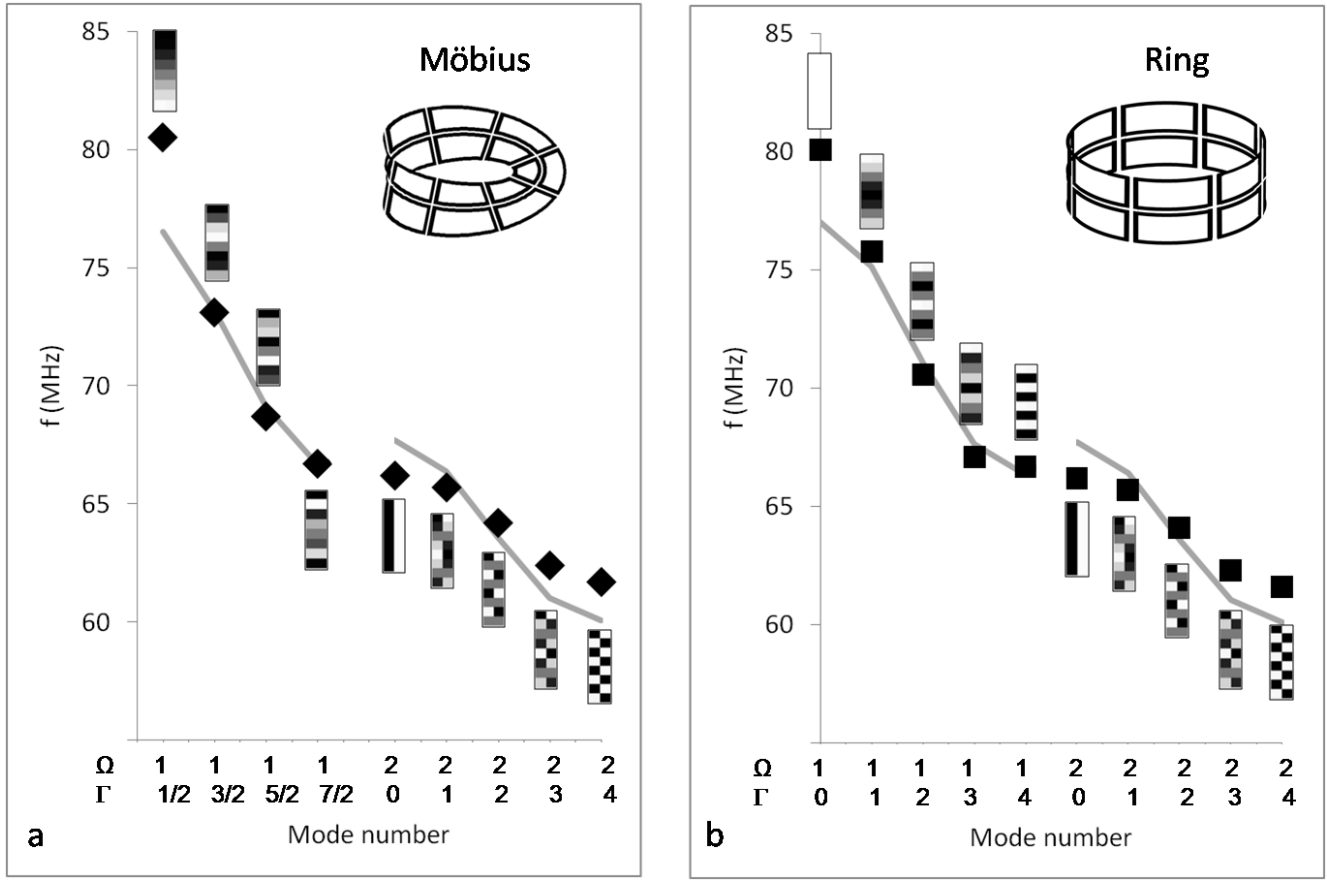

**Figure 3:** Eigenvalues and eigenfunctions for a 2 × 8 high-pass Möbius **(a)** and ring resonator **(b)**. Theoretical eigenvalues are given by gray lines, and experimental eigenvalues by symbols. In addition, for each data point the theoretical eigenfunction is displayed by rendering the current amplitudes for the 16 elements in grayscale. The figure insets show schematic drawings of each configuration; the elements interact via inductive coupling to nearest neighbors only. Model parameters for the numerical computation of theoretical eigenvalues for both resonators were obtained by measurements of the long- and short-sided inductive coupling constants and the single element resonance frequency.

**List of References**